\documentclass[reprint,superscriptaddress,notitlepage,a4paper]{revtex4-1}
\usepackage[hmargin=1.5cm,vmargin=2.0cm]{geometry}
\usepackage{float}
\usepackage{graphicx}
\usepackage{siunitx}
\usepackage{xcolor}
\usepackage[normalem]{ulem}
\usepackage{mathtools}
\usepackage{bm}
\usepackage{amsfonts}

\newcommand{\textsub}[2]{{#1}_{\text{#2}}}

\newcommand{\FIG}[1]{Fig.\,\ref{#1}}

\begin{document}
\title{Accelerating Simulated Annealing of Glassy Materials with Data Assimilation}
\author{Yuansheng Zhao}
\affiliation{Department of Physics, the University of Tokyo, Tokyo, Japan}
\affiliation{Institute for Solid State Physics, the University of Tokyo, Chiba, Japan}

\author{Ryuhei Sato}
\affiliation{Department of Physics, the University of Tokyo, Tokyo, Japan}
\affiliation{Advanced Institute for Materials Research, Tohoku University, Miyagi, Japan}

\author{Shinji Tsuneyuki}
\affiliation{Department of Physics, the University of Tokyo, Tokyo, Japan}

\begin{abstract}
The ultra-long relaxation time of glass transition makes it difficult to construct atomic models of amorphous materials by conventional methods. We propose a novel method for building such atomic models using data assimilation method by simulated annealing with an accurately computed
interatomic potential augmented by penalty from experimental data. The advantage of this method is that not only can it reproduce experimental data as the structure refinement methods like reverse Monte Carlo but also obtain the reasonable structure in terms of interatomic potential energy. In addition, thanks to the interatomic potential, we do not need high $Q$ range diffraction data, which is necessary to take into account the short-range order. Persistent homology analysis shows that the amorphous ice obtained by the new method is indeed more ordered at intermediate range.
\end{abstract}
\maketitle

Despite lacking long ranged order, glassy materials still possess short and intermediate ranged order, which is important for understanding the nature of glass transition \cite{adam1965temperature,kirkpatrick1989scaling,bouchaud2004adam,banerjee2014role}, anomalous vibrational properties \cite{taraskin2001origin,shintani2008universal,malinovsky1986nature} and for application of amorphous materials. As temperature decreases, the relaxation time of supercooled liquid increases drastically and approaches to \SI{e2}{s} order near the glass transition temperature. Such slow dynamics apparently cannot be handled by conventional computer simulation, as a consequence, the simulated structures can be under-relaxed, which is one of the biggest obstacles for obtaining good atomic models for these glassy materials and studying their structures.
Several advanced techniques have been proposed to overcome this issue using Monte Carlo simulations with swap update or umbrella sampling \cite{ninarello2017models,gutierrez2015static,torrie1977nonphysical}. Although these methods are efficient, it is often inappropriate to apply these methods to obtain the structures of experimental glassy materials due to special requirements of the methods (poly-dispersive system, good bias potential, etc.).
On the other hand, with the development of experimental techniques, x-ray or neutron diffraction data of glassy materials have been obtained to very high quality \cite{narten1972diffraction,kohara2003high,okuno2005structure}, even for small molecular glasses that crystalize easily \cite{mizuno2022intermolecular}. The structure factor $S(Q)$ obtained by normalizing corrected scattering intensity of such experiments is directly related to the radial distribution function (RDF) via Fourier transform, and there are several techniques for refining structure to match experiment data. The most widely used one is the reverse Monte Carlo (RMC) method \cite{mcgreevy1988reverse}, which tries to minimize the squared difference of structure factor between simulation box and experiment. Potential iteration refinement is another kind of methods, in which the effective potential that reproduces the experimental data is obtained by either Boltzmann relation \cite{soper1996empirical,soper2005partial,soper2012computer}, inverse reinforcement learning \cite{zhao2021structural} or other methods. 

For liquid state, the structure refinement methods work rather well if suitably implemented. However, for glassy states, as the system cannot explore the whole configurational space, it is vital to use a good initial condition for structural refinement to avoid producing significant artifacts. Yet, the difficulty for obtaining good configurations using conventional simulation methods puts us in a dilemma of dead loop.

In order to tackle the problems of glassy materials, it is natural to consider using diffraction data to accelerate the simulation itself rather than to refine the structures built by simulation. Here, we propose that when good potential $E$ is available for the system under study, acceleration can be achieved using data assimilation method by augmenting $E$ with a penalty function from experimental data \cite{tsujimoto2018crystal}, i.e.
\begin{equation}
    F(\bm R):=E(\bm R)+\lambda N D[\textsub IB(\bm R, Q),\textsub IE(Q)],
\end{equation}
where $\bm R$ and $N$ are the coordinates and number of all atoms; $D$ is a penalty function of simulated diffraction intensity $\textsub IB(\bm R, Q)$ and experimental one  $\textsub IE(Q)$, and $\lambda$ is a constant determining the strength of $D$. The penalty $D$ is defined so that it is minimized when $\textsub IB(\bm R, Q)\equiv c\textsub IE(Q)$ with $c$ a constant, i.e. the shape of simulated diffraction pattern totally agrees with that of the experiment. Both $E$ and $D$ should have local or global minimum at around the structure obtained in the experiment. In interatomic potential, local minimum corresponds to the ordered structure according to the short-range interaction between atoms, whereas diffraction pattern mainly reflects intermediate or long-range order of the structure as it is predicted from $Q$ values in diffraction pattern. Therefore, except the common one around the structure obtained in the experiment, local minimums in $E$ and $D$ are different. The local minimums in $E$ and $D$ are expected to be canceled out, if $E$ and $D$ are combined appropriately to form $F$, which accelerates the structure search. We tested previously that for crystalline phase, the correct structure can be obtained directly by simulation using $F$ but not $E$ \cite{tsujimoto2018crystal,Yoshikawa, adachi2019search}. It is expected that for glassy materials, this method helps crossing the barrier of long relaxation time and can accelerate simulation to obtain lower energy and better agreement with experiment.

Before going further, we would like to clarify the difference between current data assimilation method and structure refinement methods. 
The idea of jointly minimizing $D$ and $E$ is not unique to the current data assimilation method because some structural refinements such as hybrid RMC \cite{opletal2002hybrid,petersen2003structural} also incorporate both $D$ and $E$.
However, in these schemes, the initial structure is prepared by conventional methods (e.g. simulated annealing) and atoms are then moved to match experiment data (possibly also taking into account of constraints or $E$), which does not guarantee that the obtained structure is energetically favored (it will actually raise the energy in many cases). In contrast, the data assimilation generates glass structures directly from liquid state by simulated annealing with interatomic potential energy combined with experimental penalty. In this simulation, interatomic potential and experimental penalty decrease simultaneously during the simulated annealing. As we will show later, this treatment can obtain lower energy structure alongside better agreement with experiment, and thus gains much higher reliability. In addition, as the interatomic potential can easily handle short ranged order, only low $Q$ region of the diffraction data is needed for the simulation.

In this study, the penalty $D$ is defined using the correlation coefficient between $\textsub IB$ and $\textsub IE$:
\begin{equation}
	D=1-\frac{\overline{(\textsub{I}{B}(\bm R,Q)-\textsub{\bar I}{B})(\textsub{I}{E}(Q)- \textsub{\bar I}{E})}}{\sqrt{\overline{(\textsub{I}{B}(\bm R,Q)-\textsub{\bar I}{B})^2}\cdot\overline{(\textsub{I}{E}(Q)- \textsub{\bar I}{E})^2}}},
\end{equation}
where the bar $\bar I:=\frac{1}{\textsub Q{max}-\textsub Q{min}}\int_{\textsub Q{min}}^{\textsub Q{max}} I(Q) \,\mathrm dQ$ denotes $Q$-average of $I$. This form of $D$ can capture both the height and position of each peaks, and can also handle experimental data that cannot be normalized to structure factor $S(Q)$ because of lacking high $Q$ part or large noise. We tested that the RMC-styled $D:=\overline{(\textsub{S}{B}(\bm R,Q)-\textsub{S}{E}(Q))^2}$ also works well, even better than correction coefficient sometimes \footnote{See Fig.\,1(c) in supplementary materials}.
The force coming from $D$ can be computed by differentiation of $D$. Because the intensity $I(Q)$ or RDF $g(r)$ are two-body functions, at any fixed time, the force is pairwise additive, which makes the implementation fairly straight forward. For classical simulations, the computation of scattering intensity and force can become a bottleneck as communication over processes and numerical integration are needed. We make the following optimizations: first, the pair force is only calculated on the grid for tallying RDF $g(r)$ (used to evaluate $I(Q)$) and interpolation is used for other $r$;
and second, as $\textsub IB(Q)$ does not change significantly over short time, this force table is only updated every 5 steps. Using these optimizations, the computation times involving $D$ is very short compared with $E$. We added self-written code \footnote{The code is available at https://github.com/YuanshengZhao/imp} for computing the force from the penalty function $D$ into the LAMMPS package \cite{thompson2022lammps} for molecular dynamics simulation in this study.

\begin{figure}
    \centering\includegraphics[width=\linewidth]{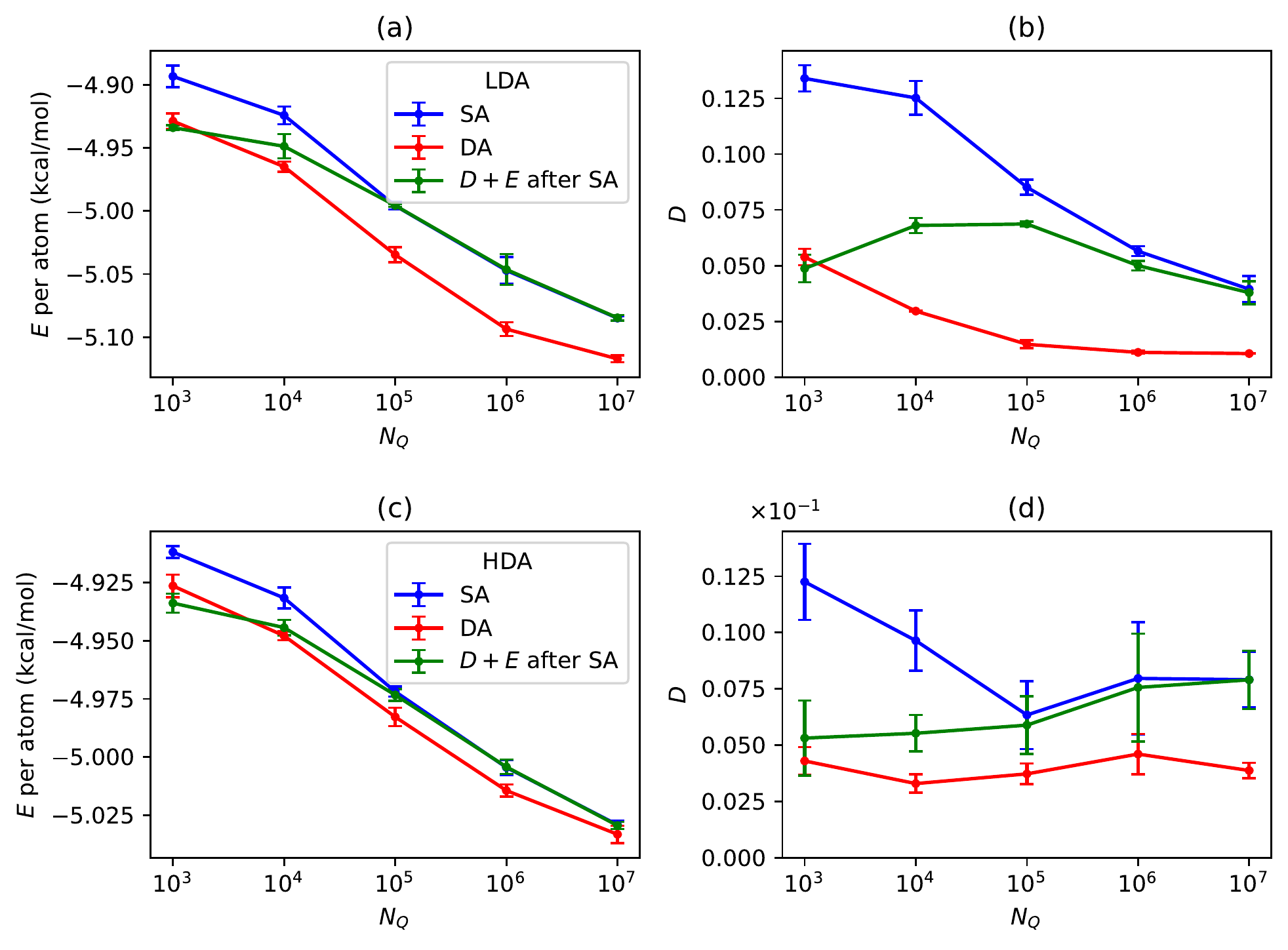}
    \caption{\textbf{(a)} Interatomic energy $E$ of the LDA structures obtained with simulated annealing with $E$ (SA), data assimilation (DA, $F = E + \lambda N D$) and running with $F$ only at low temperature after the simulated annealing with $E$, for different cooling steps $N_Q$. The error bar is the standard deviation of 3 independent runs. \textbf{(b)} Same as (a), but for penalty from experimental data $D$, 0 is absolute agreement. Data assimilation consistently gives lower energy and better agreement with experiment, but this cannot be achieved by using $F$ only at low temperature after the simulated annealing with $E$. \textbf{(c), (d)} Same as (a) and (b), for HDA. While simulated annealing can already produce fairly good structures, data assimilation can still boost the performance.}\label{fg:nq_all}
\end{figure}

Using the new method, we compare the energy and structures of low density (LDA) and high density (HDA) amorphous ice obtained by simulated annealing and data assimilation. The TIP4P/ice model \cite{abascal2005potential} which is optimized for simulation for solid state, is used for evaluating the interatomic potential energy $E$. The force is cut off at \SI{12}{\angstrom} for both $D$ and $E$ and PPPM method \cite{hockney2021computer} is used for long range Coulomb potential. We use x-ray diffraction data \cite{bizid1987structure} between 0.7 and \SI{6}{\angstrom^{-1}} to calculate the penalty \footnote{Because the electronic density of water is almost spherical, the x-ray structure factor is approximately the Fourier transform of RDF of molecular centers. As the virtual site M of TIP4P model is the closest to the geometric center of electronic density, the $\textsub g {MM}(r)$ is used to calculate $S(Q)$ instead of $\textsub g {OO}(r)$. Using $\textsub g {OO}(r)$ will not change the qualitative result.}. The simulation procedure is as follows. The liquid structure containing 1000 rigid water molecules with the same density as glass (\SI{0.94}{g/cm^3} for LDA and \SI{1.17}{g/cm^3} for HDA) equilibrated using \SI{e2}{ps} simulation at \SI{323}{K} is used as initial condition. We then run simulated annealing at constant volume with $F$ (data assimilation) or $E$ (conventional simulated annealing). The initial temperature is \SI{323}{K} and the temperature is decreased to \SI{77}{K} linearly at the end of the simulated annealing. After this simulated annealing procedure, the experimental penalty $D$ is turned off and \SI{30}{ps} simulation is performed for thermalization with only $E$.
Finally, another \SI{10}{ps} simulation is conducted to collect the structural data. The time step is fixed at \SI{1}{fs} and the temperature is controlled using Nos\'e--Hoover chain method \cite{martyna1992nose}.

\FIG{fg:nq_all}(a)(b) shows the interatomic energy $E$ and experiment penalty $D$ of the structures obtained with normal simulated annealing and data assimilation for cooling time \SI{1}{ps} $\sim$ \SI{10}{ns} (\SI{2.46e14}{K/s} $\sim$ \SI{2.46e10}{K/s}) for LDA.
The value for $D$ strength parameter $\lambda$ is set to \SI{3}{kcal/mol}, which gives the lowest energy from fast test runs (1 $\sim$ \SI{10}{kcal/mol} all produce very low energy)  \footnote{See Fig.\,1(a)(b) in supplementary materials}. First we observe that for conventional simulated annealing, the $D$ gradually decreases as cooling step, $N_Q$, increases. This strongly suggests that the global minimums of $D$ and $E$ 
coincide, which is a key for the success of data assimilation method. A direct comparison reveals that with data assimilation, not only can we consistently get better agreement with experiment, but also lower energy $E$, showing that adding the experimental penalty is indeed helpful to reduce local minimums of $E$ and can achieve huge acceleration (around $10\times$ when comparing $E$ for slow cooling). We note that running simulation with $F$ for $10^5$ steps only at low temperature from structures obtained by conventional simulated annealing -- a kind of structure refinement -- cannot achieve similar effect (see green lines in \FIG{fg:nq_all}), thus it is concluded that the cooling process itself is also a key to success and structural rearrangement is difficult even with the introduction of $D$ at low temperature. \FIG{fg:nq_all}(c)(d) shows the results of the same numerical experiment for HDA. It is found that the result is qualitatively the same, albeit the effect of data assimilation is weaker because conventional simulated annealing can already get structures with low $E$ and $D$. The weaker effect of data assimilation to HDA is explained by the structural difference between HDA and LDA. The first sharp diffraction peak (FSDP) of HDA is located at much higher $Q$ than LDA, indicating that the characteristic structural length of HDA is short. It is also argued from experimental data that locally, LDA is similar to crystal ice while HDA is similar to liquid water \cite{finney2002structures}. The structural difference shows that HDA is inherently less ordered than LDA and building atomic model of HDA is easier.

\begin{figure}
    \centering\includegraphics[width=\linewidth]{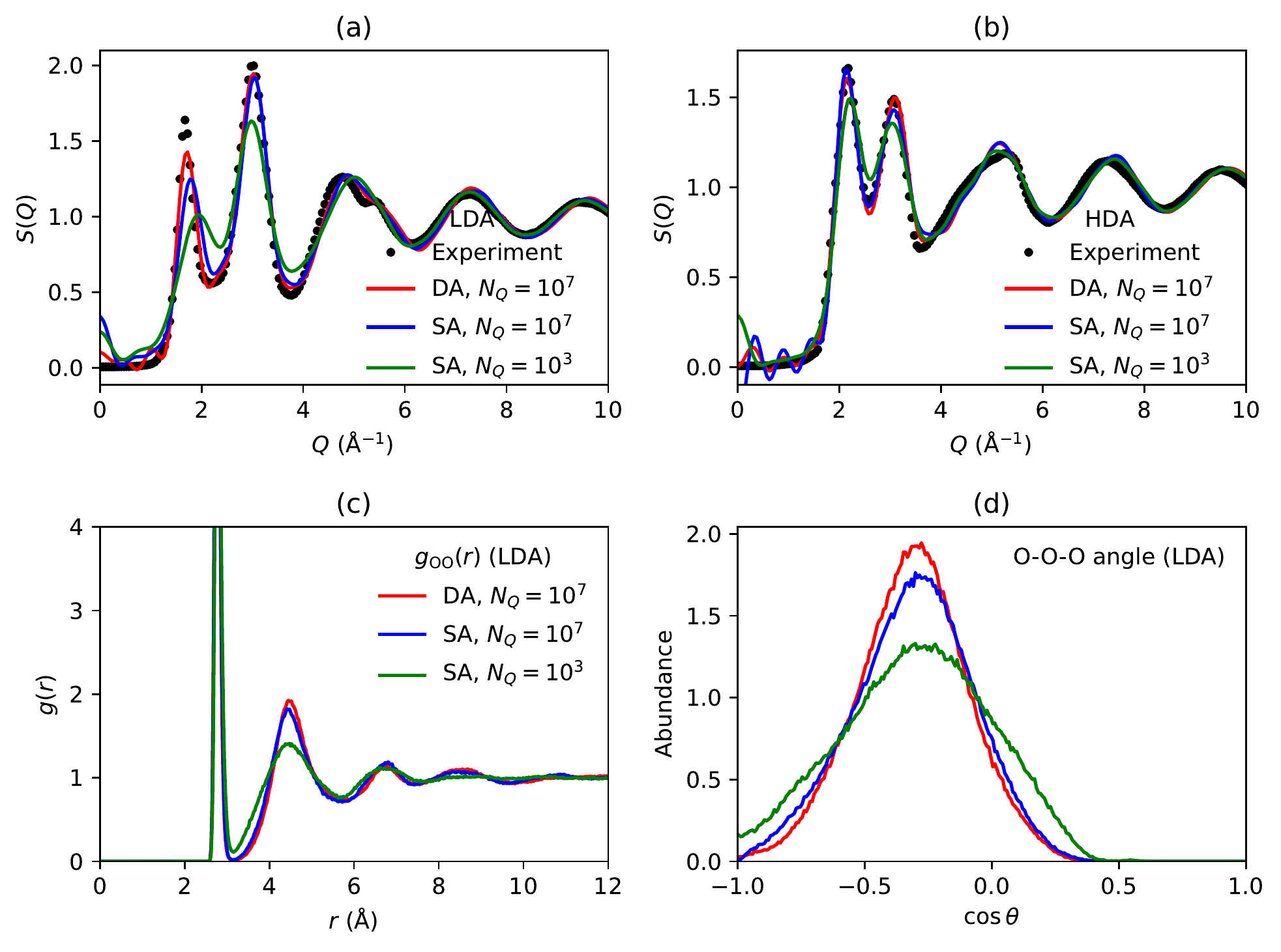}
    \caption{\textbf{(a)} x-ray structure factor $S(Q)$ of LDA, obtained by experiment, simulated annealing (SA) and data assimilation (DA). Slower cooling produces higher first peak at lower $Q$, but only data assimilation reproduces the correct position. \textbf{(b)} Same as (a), for HDA. The structural difference is smaller. \textbf{(c)} Difference of $\textsub g{OO}(r)$ of LDA. Data assimilation produces the most ordered structure. \textbf{(d)} Same as (c) for O--O--O angle distribution.}\label{fg:sq}
\end{figure}

\FIG{fg:sq} compares the simulated x-ray structure factor $S(Q)$. 
For LDA, it is found that slower cooling produces higher FSDP but is insufficient to reproduce the experimental data as its position is located at lower $Q$. However, the FSDP obtained with data assimilation has the correct $Q$ position and the largest amplitude, indicating that the intermediate ranged order is more developed.  With data assimilation, the shoulder near \SI{5}{\angstrom^{-1}} is also more pronounced. 
In real space, for structures made by data assimilation, the radial distribution of oxygen atoms shows more pronounced second peak, whose position corresponds to the position of FSDP. 
The angle distribution of non-directly-bonded O--O--O is also sharper.
This clearly reveals that the structures built using data assimilation are more ordered at intermediate range and hence have lower energy.
On the other hand, the structural difference of HDA is much smaller, consistent with smaller energy difference \footnote{See Fig.\,2 in supplementary material for comparison.}.

\begin{figure}
    \centering\includegraphics[width=\linewidth]{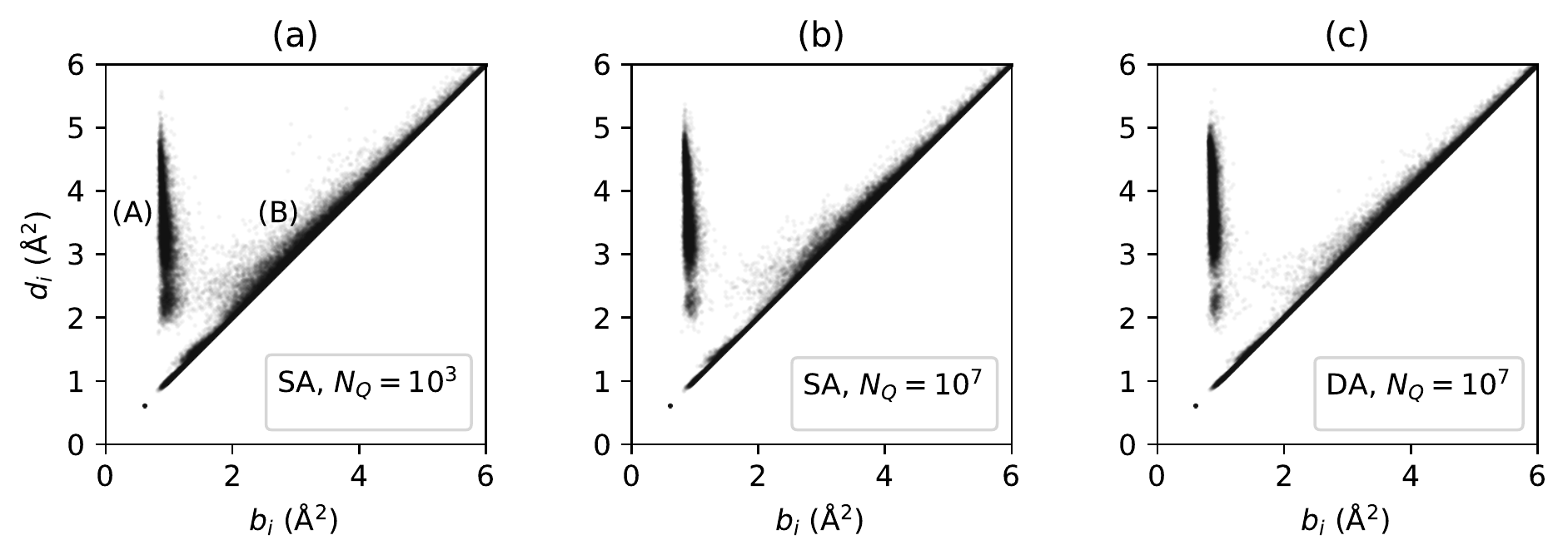}
    \caption{First order persistent diagram $D_1$ of structures produced by simulated annealing with fast cooling \textbf{(a)}, slow cooling \textbf{(b)} and data assimilation with slow cooling \textbf{(c)}. Different atomic types are not distinguished. The density in region (B) is reduced by slower cooling or using data assimilation.}\label{fg:ph}
\end{figure}
\begin{figure}
    \centering\includegraphics[width=\linewidth]{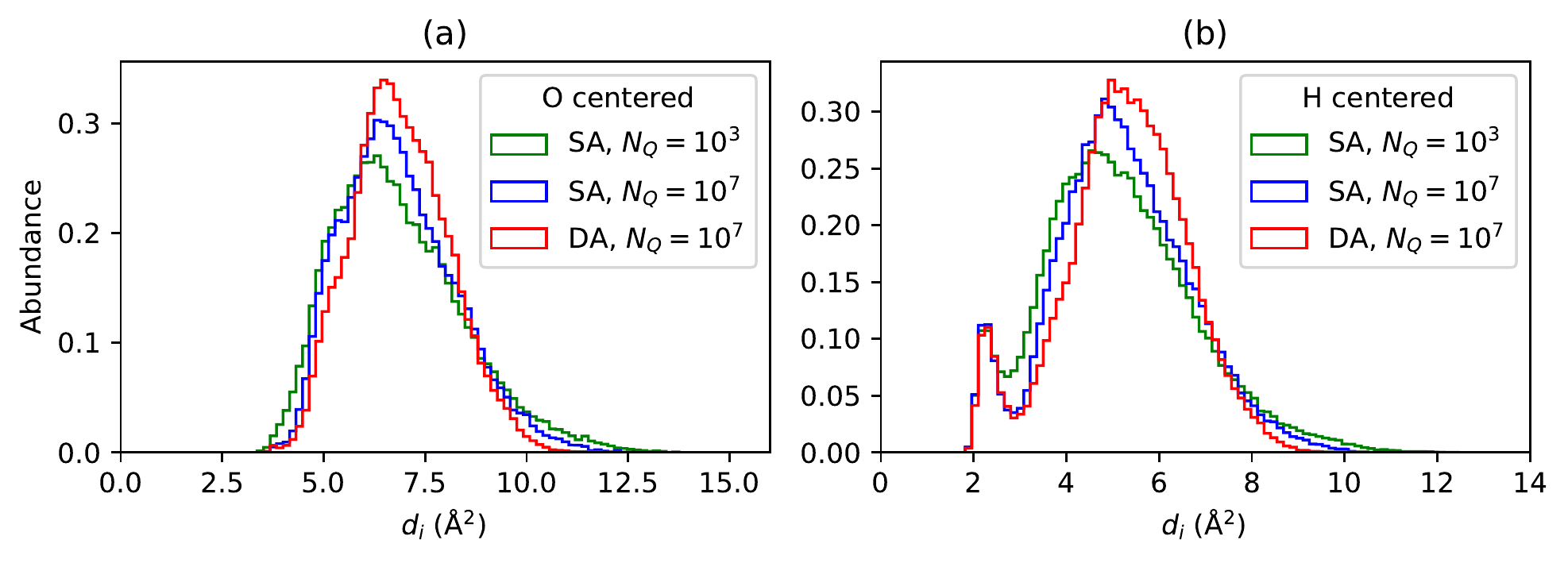}
    \caption{Histogram of $d_k$ of second order persistence $D_2$, i.e. size of all cavities of the structures, when including only O atoms \textbf{(a)} and H atoms \textbf{(b)}. In both cases, with data assimilation, the distribution is sharper and there are less large cavities, indicating more ordered structure.}\label{fg:phdi}
\end{figure}

We further analyzed the topological features of the structures using persistent homology \cite{hiraoka2016hierarchical}. The input for this analysis is the coordinates of the atoms, and we gradually increase the ``atomic radius'' $r$ from 0. For the first order persistent homology $D_1$, consider each ring $c_k$ consisting of atoms in the structure, we note that there is a value $b_k$ such that the ring first appears when $r=\sqrt{b_k}$ (``birth'', atoms start to touch adjacent ones on all edges), as well as another value $d_k\ge b_k$ such that the ring disappears when $r=\sqrt{d_k}$ (``death'', all space inside the ring gets filled). By definition, $b_k$ and $d_k$ is related to the edge length and overall size of the rings.
Similarly, cavities instead of rings can be considered, and this is the second order persistent homology $D_2$. The CGAL \footnote{CGAL: www.cgal.org/} and PHAT \footnote{PHAT: https://bitbucket.org/phat-code/phat} libraries are used to compute persistent homology in this study.

\FIG{fg:ph} shows the first order persistence diagram $D_1$ which is plotted by putting all the persistence pairs $(b_k,d_k)$ on $\mathbb R^2$ plane. All H and O atoms are included without distinguishing. There are two distinct features: a vertical band (A) and a $45^\circ$ band located slightly above the diagonal (B). From the value of $b_k$, only (A) represents hydrogen bonded network which is energetically favorable.
The difference of the persistence diagram shows that using data assimilation is equivalent to slower cooling, with the effect of reducing the density in (B) region
and thus lower energy. 
From the distribution of $d_k$ of $D_2$ (i.e. size of cavities in the structure) as shown in \FIG{fg:phdi}
\footnote{$D_2$ persistence diagram is shown in Fig.\,3 in supplementary materials.}, we see that data assimilation generates more uniform cavity sizes (sharper distribution). Importantly, very large cavities which represents defects in the structure are largely suppressed when introducing data assimilation in the simulation.
Thus, we conclude that data assimilation indeed produced structure which is more ordered at intermediate range.

As HDA is experimentally produced by compressing ordinary ice instead of liquid quenching at high pressure, it is interesting to study the pressure induced amorphization using the current method. Starting from cubic ice Ic with random hydrogen orientation, the structure is gradually compressed to \SI{1.30}{g/cm^3} and then expanded to HDA density (\SI{1.17}{g/cm^3}) by modifying the edge length of the simulation box. When running compression/expansion at \SI{140}{K} (close to glass transition temperature), using $F$ instead of $E$ also results in better agreement with experiment and lower energy $E$ if the compression/expansion steps are greater than $5\cdot 10^4/5\cdot 10^4$ so that there is sufficient time for the structure to finish the amorphization. We finally emulated the transition from HDA to LDA by gradually expanding the HDA structure to LDA density (\SI{0.94}{g/cm^3}) at \SI{140}{K}, and obtained qualitatively the same result. It is worth noting that the LDA obtained from HDA using data assimilation has consistently higher energy and lower FSDP than that obtained from liquid \footnote{See Fig.\,4 in supplementary materials}. This suggests that even with the acceleration of data assimilation, it is difficult to reach the state with lowest energy for transition from HDA to LDA. This supports the previous finding that transition from HDA to LDA is gradual with many metastable states and how much the structure can relax determines the final energy and intermediate ranged structure \cite{guillot2003polyamorphism,winkel2009relaxation}.

From the numerical experiments with poly-amorphism of ice, we showed that augmenting penalty function $D$ to the interatomic potential $E$ can indeed accelerate the simulated annealing process and build better atomic model of glassy materials. As a lower energy can be obtained, the data assimilation has higher reliability than structural refinement methods for thermally arrested glassy state when accurately computing $E$ is possible. Because the final structure is maintained using only $E$, it is also suitable for studying the dynamic properties theoretically.

This work was supported by Japan Society for the Promotion of Science (JSPS) KAKENHI Grant-in-Aid for Scientific Research on Innovative Areas ``Hydrogenomics'', No. JP18H05519. Y.Z greatly appreciates the financial support by JSPS Research Fellowship for Young Scientists.

\bibliography{rc.bib}

\newpage
\appendix
\section*{Supplementary material}
\setcounter{figure}{0}
\begin{figure}[H]
    \centering\includegraphics[width=\linewidth]{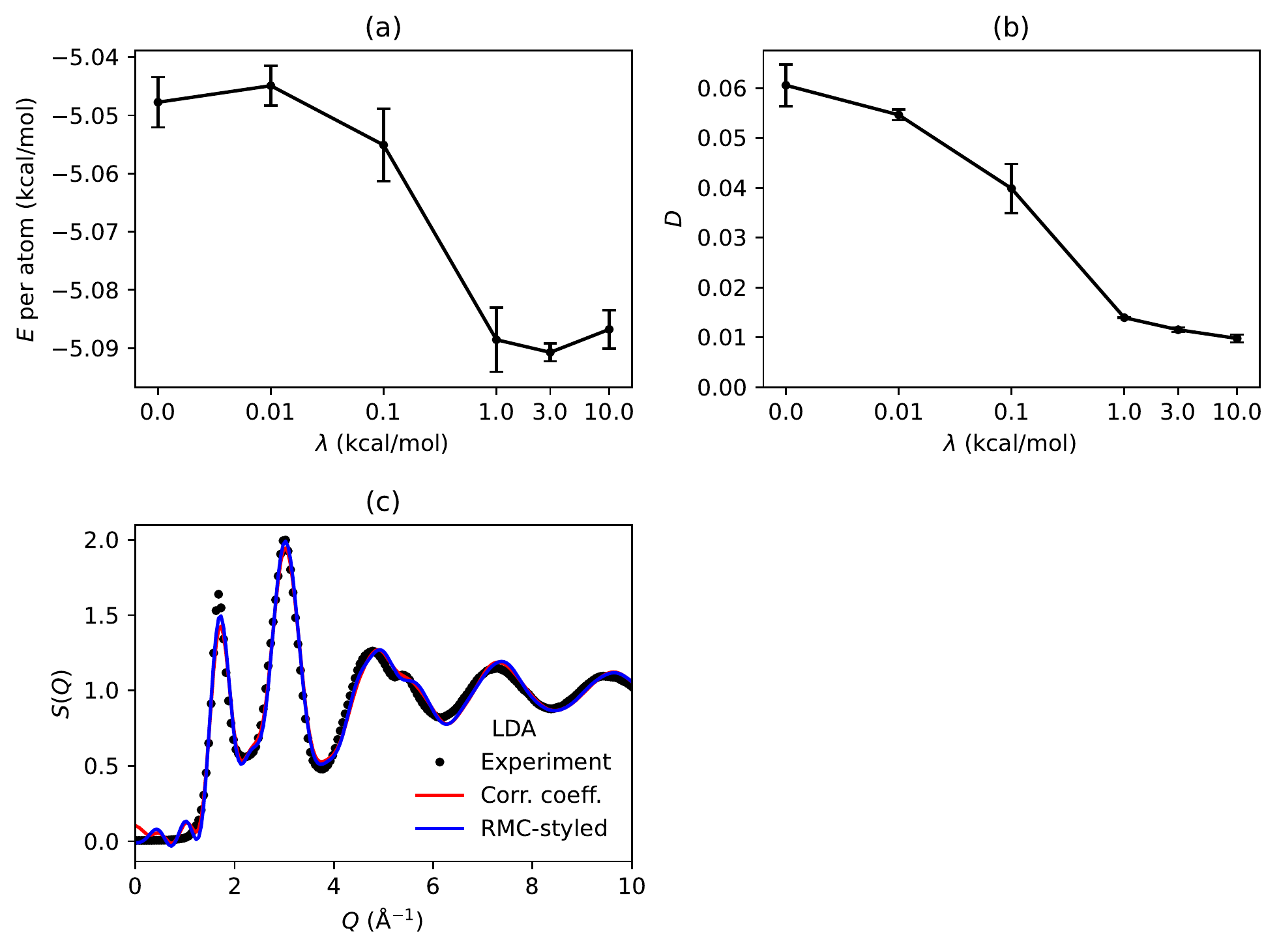}
    \caption{\textbf{(a)} $\lambda$ (strength of penalty $D$) dependence of final energy $E$ for simulated annealing with LDA in $10^6$ steps. We get low energy for $1\le\lambda\le\SI{10}{kcal/mol}$. \textbf{(b)} Same as (a), for final penalty $D$. \textbf{(c)} $S(Q)$ obtained by data assimilation with RMC-styled penalty $D:=\overline{(\textsub{S}{B}(\bm R,Q)-\textsub{S}{E}(Q))^2}$, the $Q$ range used is $[0.7,\SI{4}{\angstrom^{-1}}]$ and $\lambda=\SI{9}{kcal/mol}$. We see that this $D$ can also get very good result, even better with $D$ defined by correction coefficient, presumably because there is no arbitrary scaling constant in $S(Q)$ as in $I(Q)$.}\label{fg:lmd}
\end{figure}

\begin{figure}[H]
    \centering\includegraphics[width=\linewidth]{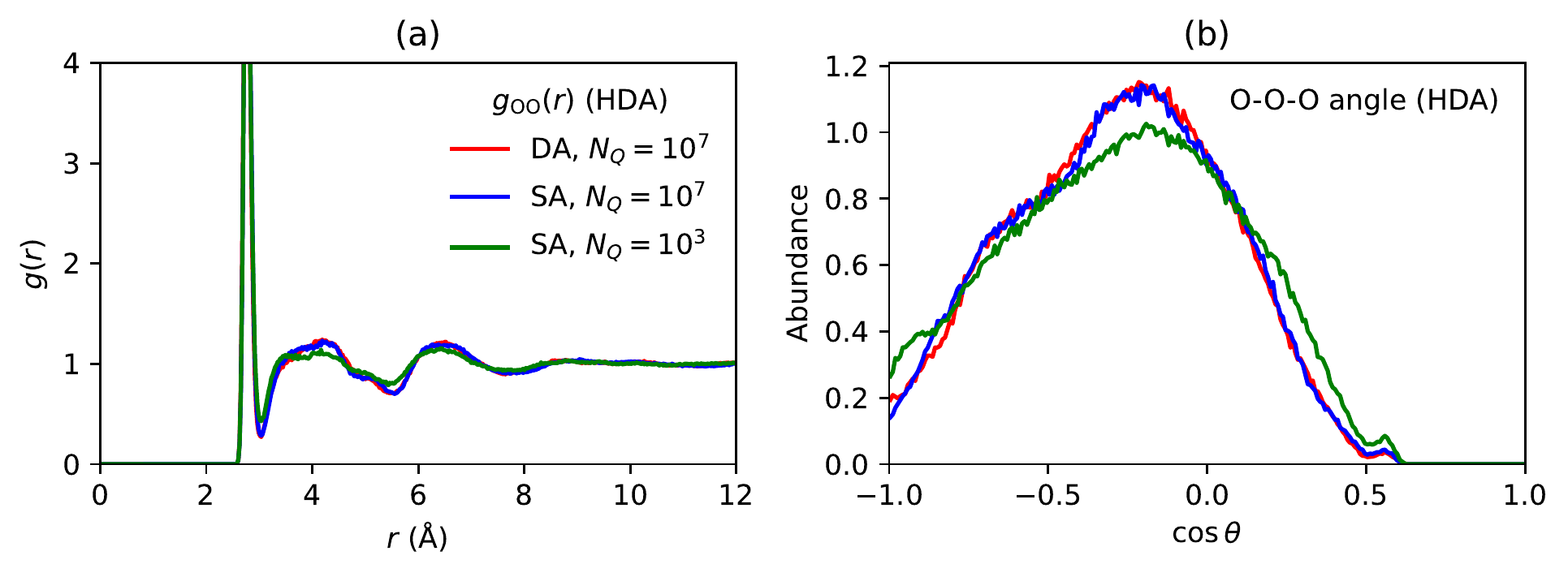}
    \caption{The $\textsub{g}{OO}(r)$ \textbf{(a)} and O--O--O angle distribution \textbf{(b)} of HDA obtained by simulated annealing (SA) and data assimilation (DA). The difference is small compared with LDA as expected from small difference in $S(Q)$.}\label{fg:grhda}
\end{figure}

\begin{figure}[H]
    \centering\includegraphics[width=\linewidth]{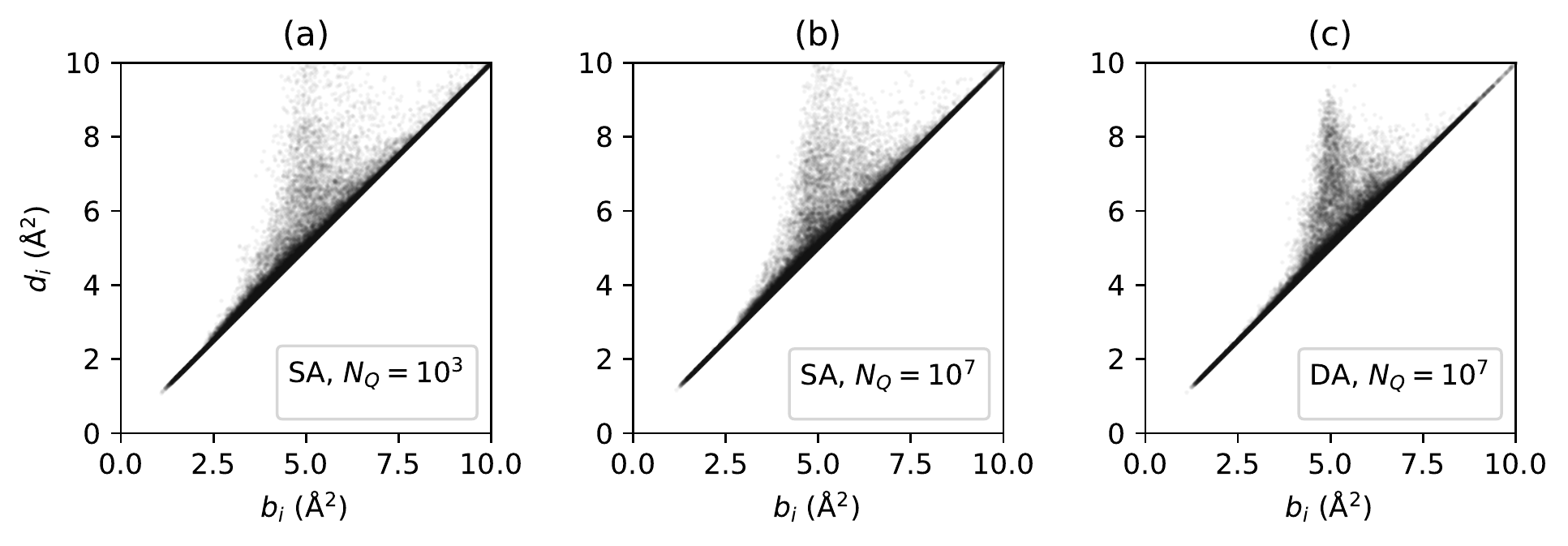}
    \caption{Second order persistent diagram $D_2$ of LDA structures produced by simulated annealing with fast cooling \textbf{(a)}, slow cooling \textbf{(b)} and data assimilation with slow cooling \textbf{(c)}. Different atomic types are not distinguished. This also shows that data assimilation produces more ordered structures.}\label{fg:ph2}
\end{figure}

\begin{figure}[H]
    \centering\includegraphics[width=\linewidth]{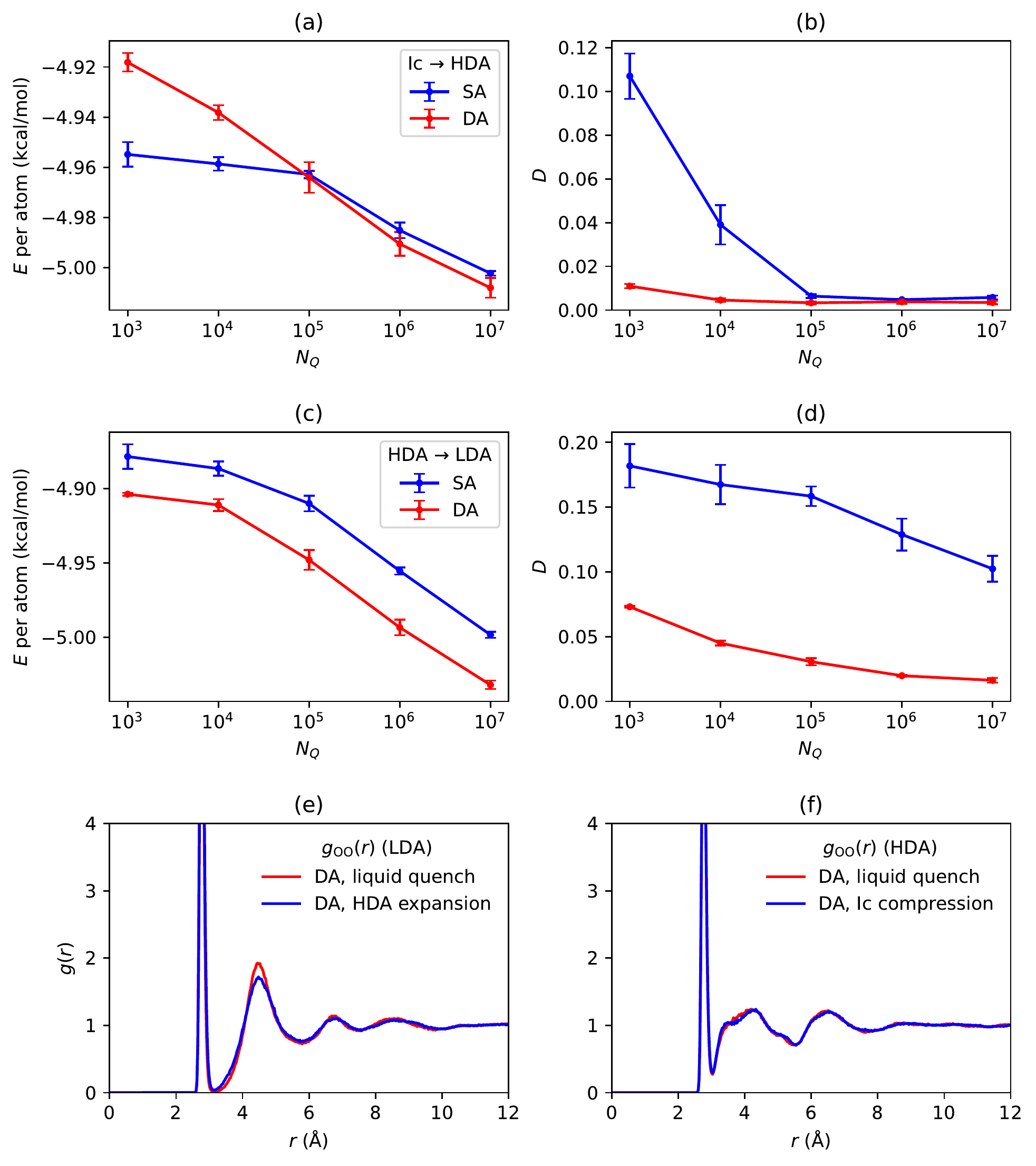}
    \caption{\textbf{(a)} Compression speed dependence of final energy $E$ of HDA obtained from compressing Ic instead of simulated annealing from liquid. The $N_Q$ here is the total steps for compression (to \SI{1.30}{g/cm^3}) and expansion (to \SI{1.17}{g/cm^3}), with each process taking up $N_Q/2$ steps. The temperature is fixed at \SI{140}{K} (close to glass transition temperature) during compression and expansion and directly switched to \SI{77}{K} for data collection. It shows that data assimilation can also lower the energy in this case except for very quick compression when amorphization is insufficient without data assimilation. \textbf{(b)} same as (a), for penalty. \textbf{(c)} Expansion speed dependence of $E$ for simulated expansion from HDA to LDA at \SI{140}{K}. The temperature is directly switched to \SI{77}{K} for data collection. The effect of data assimilation is the same as before. The final energy is higher than the case of simulated annealing from liquid. \textbf{(d)} Same as (c) for penalty. \textbf{(e)} Comparison of O--O $g(r)$ produced by simulated annealing from liquid and expansion of HDA. The former can relax more and is more ordered, supporting the view that the that transition from HDA to LDA is gradual with many metastable states and how much the structure can relax determines the final energy and intermediate ranged structure. \textbf{(f)} Comparison of O--O $g(r)$ from simulated annealing from liquid and compression-expansion of Ic. We may draw a similar conclusion as LDA case.}\label{fg:nqcompress}
\end{figure}

\end{document}